\documentclass[useAMS,usegraphicx,usenatbib]{mn2e}                
\usepackage{times}

\newcommand{\Msun}{\ensuremath{\,{\rm M}_\odot}}                  
\newcommand{\Rsun}{\ensuremath{\,{\rm R}_\odot}}                  
\newcommand{\Teff}{\ensuremath{T_{\rm eff}}}                      
\newcommand{\Mjup}{\ensuremath{\,{\rm M}_{\rm Jup}}}              
\newcommand{\Rjup}{\ensuremath{\,{\rm R}_{\rm Jup}}}              
\newcommand{\Teq}{\ensuremath{T_{\rm eq}^{\,\prime}}}             
\newcommand{\safronov}{\ensuremath{\Theta}}                       
\newcommand{\ms}{\,m\,s$^{-1}$}                                   
\newcommand{\mss}{\,m\,s$^{-2}$}                                  
\newcommand{\as}{\ensuremath{^{\prime\prime}}}                    
\newcommand{\am}{\ensuremath{^\prime}}                            
\newcommand{\FeH}{\ensuremath{\left[\frac{\rm Fe}{\rm H}\right]}} 
\newcommand{\pjup}{\ensuremath{\,\rho_{\rm Jup}}}                 
\newcommand{\psun}{\ensuremath{\,\rho_\odot}}                     
\newcommand{\mc}[1]{\multicolumn{2}{c}{#1}}
\newcommand{\mcc}[1]{\multicolumn{3}{c}{#1}}

\newcommand{\erc}[3]{\mc{\ensuremath{#1^{+#2}_{-#3}}}}

\newcommand{\ermcc}[5]{\mcc{\ensuremath{{#1\,^{+#2}_{-#3}}\,^{+#4}_{-#5}}}}

\title[Transits in the WASP-50 planetary system]
{An extremely high photometric precision in ground-based observations of two transits in the WASP-50 planetary system}

\author[Tregloan-Reed et al.]
       {Jeremy Tregloan-Reed\,$^{1}$\thanks{Email: j.j.tregloan-reed@keele.ac.uk}, John Southworth\,$^{1}$ \\
        $^{1}$\,Astrophysics Group, Keele University, Staffordshire, ST5 5BG, UK \\
}

\begin{document} \maketitle 

\begin{abstract}
We present photometric observations of two transits in the WASP-50 planetary system, obtained using the ESO New Technology Telescope and the defocussed-photometry technique. The rms scatters for the two datasets are 258 and 211\,ppm with a cadence of 170 to 200\,s, setting a new record for ground-based photometric observations of a point source. The data were modelled and fitted using the \textsc{prism} and \textsc{gemc} codes, and the physical properties of the system calculated. We find the mass and radius of the hot star to be $0.861\pm 0.057\Msun$ and $0.855\pm0.019\Rsun$, respectively. For the planet we find a mass of $1.437\pm 0.068\Mjup$, a radius of $1.138\pm0.026\Rjup$ and a density of $0.911\pm0.033\pjup$. These values are consistent with but more precise than those found in the literature. We also obtain a new orbital ephemeris for the system: $ T_0 = {\rm BJD/TDB} \,\, 2\,455\,558.61237 (20) \, + \, 1.9550938 (13) \times E $.
\end{abstract}


\section{Introduction}                                                                                                              
\label{sec:intro}

\citet{Mayor95} discovered the first planet orbiting a normal star outside our own solar system, 51\,Peg, from radial velocity (RV) measurements. The first transiting planet was found using photometric observations of a system, HD\,209458, already known from RV measurements to host a planet \citep{Henry+00apj,Charbonneau+00apj}. OGLE-TR-56 was the first planet discovered from its transits \citep{Udalski+02aca3,Konacki2003}. The transit detection method uses photometry to measure the change in received flux from a star when a planet crosses the stellar disc. Since then, ground-based transit detection surveys such as WASP \citep{SuperWasp} and HAT \citep{HAT} have been set up around the world. Once a survey has discovered a transiting extrasolar planet (TEP), follow-up spectroscopic and photometric observations are required to properly characterise the system. 

To achieve high-precision photometry requires not only a high signal-to-noise ratio (S/N) but also nullification of many systematic effects inherent in ground-based photometry. \citet{Sou2009} investigated the use of defocussed telescopes to obtain high-precision photometry. Telescope defocussing causes the light from point sources to be distributed over thousands of CCD pixels. This allows the use of longer exposure times, which means that the CCD is read out less often. This reduced dead time means that more photons can be collected over a given time interval, leading to lower Poisson and scintillation noise. Flat-fielding noise also averages down by an order of magnitude due to the large number of pixels in the software aperture, and changes in seeing become inconsequential. Many researchers have used the technique of defocussed photometry to obtain precise measurements of the parameters of TEP systems \citep[e.g.][]{Gillon2007,Gillon2008,Demory2007,Winn2007,Winn2007b}. We note that the defocussing technique is only suitable for bright isolated stars. If a star is too faint then the increased background and read-out noise can be detrimental. If the target and comparison stars are in a crowded field then defocussing will cause their point spread functions (PSF) to become blended with those of nearby stars.

The discovery of the TEP system WASP-50 was presented by \citet{Gillon2011}, who found it to comprise a TEP with a mass of $1.47\pm 0.09$\Mjup\ and radius of $1.15\pm 0.05$\Rjup, orbiting a cool star with mass and radius $0.89\pm 0.08$\Msun\ and $0.84\pm 0.03$\Rsun. We observed WASP-50 with the aim of improving its measured physical properties, using the telescope-defocussing approach. We used the New Technology Telescope (NTT) operated by ESO at La Silla, Chile. This telescope is alt-az mounted and has a thin primary mirror whose shape is maintained using active optics. Whilst these qualities are not outwardly well-suited to the telescope-defocussing approach, we have previously found the NTT to work well for this type of observations \citep{Hellier2011,Jeremy2012}. We observed two complete transits of WASP-50 using the NTT, achieving extremely low photometric scatters of 258 and 211 parts per million (ppm), respectively, versus a fitted model. To our knowledge the latter is the lowest scatter ever achieved in ground-based photometry per point for a point source.

Some of the highest photometric precisions previously accomplished for a TEP system are  
479\,ppm for CoRot-1 using the 8.2\,m VLT \citep{Pont2010}, 
478\,ppm for WASP-4 using the 6.5\,m Magellan Baade telescope \citep{Winn2009b}, 
470\,ppm for WASP-10 using the university of Hawaii 2.2\,m telescope \citep{Johnson2009}, 
387\,ppm for WASP-2 using a 1.5\,m telescope \citep{Sou2010b}, 
and 316\,ppm for TrES-2 using the 10.4\,m Gran Telescopio Canarias \citep{Colon2010}. 
The highest photometric precision we are aware of from a ground-based telescope was previously 258\,ppm in time-series observations of stars in the open cluster M67 \citep{Gilliland1993}. 

An alternative metric which is well-suited to direct comparison is signal to noise per unit time. We have calculated the scatter in ppm per minute of observing time for the datasets listed above. By this metric the better of our two datasets is almost exactly equal to the best one presented by \citet{Gilliland1993}, and both of our light curves are better than any previously published ground-based photometric observations of a TEP system.


\section{Observations and data reduction}
\label{sec:data}

\begin{table*} \centering
\caption{\label{tab:obslog} Log of the observations presented in this work. $N_{\rm obs}$ is
the number of observations. `Moon illum.' and 'Moon dist.' are the fractional illumination of
the Moon, and its distance from WASP-50 in degrees, at the midpoint of the transit.}
\begin{tabular}{lcccccccccc} \hline
Date & Start time & End time &$N_{\rm obs}$& Exposure & Filter & Airmass & Moon & Moon & Aperture   & Scatter \\
     &    (UT)    &   (UT)   &             & time (s) &        &         &illum.& dist.& sizes (px) & (ppm)   \\
\hline
2011/11/20 & 00:59 & 06:02 & 127 & 120--150 & Gunn $r$ & 2.62 $\to$ 1.48 & 0.384 &  91.2 & 75, 105, 120 & 258 \\
2011/11/24 & 01:08 & 06:27 & 124 & 150      & Gunn $r$ & 2.10 $\to$ 1.53 & 0.045 & 137.4 & 75, 100, 125 & 211 \\
\hline \end{tabular} \end{table*}

Two transits of WASP-50 were observed on the nights of 2011/11/20 and 2011/11/24 using the NTT with the EFOSC2 instrument operated in imaging mode. In this setup the CCD covers a field of view of $(4.1\am)^{2}$ with a plate scale of 0.12\as\,px$^{-1}$. The images were windowed down to $1100\times1600$ pixels and no binning was used, resulting in a dead time between consecutive images of 50\,s. The observations were taken through a Gunn $r$ filter (ESO filter \#784). The moon was below the horizon for half of the first transit and all of the second transit. The telescope was initially focussed and the shape of its primary mirror was adjusted to obtain the best image possible. We then applied a defocus to the telescope and performed the full observing sequence without adjusting the telescope focus or mirror shape. 

The pointing of the telescope was adjusted to allow a good comparison star to be observed simultaneously with WASP-50. The comparison star used was 2MASS J02544939$-$1051548, which is of a similar apparent magnitude and colour to WASP-50. The 2MASS $J-K_s$ colour indices of the two objects are $0.432$ for WASP-50 and $0.357$ for the comparison star \citep{2mass}. We were able to keep the telescope autoguided through all observations. An observing log is given in Table\,\ref{tab:obslog}.

\begin{figure} \includegraphics[width=0.46\textwidth,angle=0]{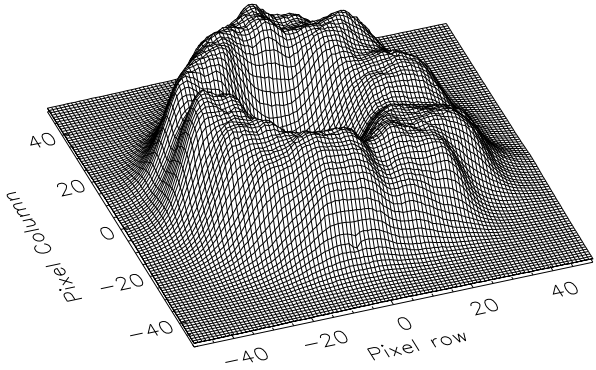} 
\caption{\label{fig:psf} Surface plot of the PSF of WASP-50 in an image taken
at random from the observing sequence on the night of 2011 October 24th. The 
x and y axes are in pixels. The lowest and highest counts are 684 and 24\,726 
ADUs, respectively, and the z axis is on a linear scale.} \end{figure}

Fig.\,\ref{fig:psf} shows the shape of the point spread function (PSF) of WASP-50 in an image taken at random from the observing sequence on the night of 2011/10/24. Fig.\,\ref{fig:psf} provides an interesting comparison with similar observations of WASP-4 and WASP-5 from a 1.5\,m telescope of more traditional design (see fig.\,1 in \citealt{Sou2009} and fig.\,1 in \citealt{Sou2009b}). The PSF for the current observations shows a much more rounded annulus of high counts, which allows a smaller amount of defocussing to be used to collect a given number of photons without saturating individual pixels.

We reduced the data in an identical fashion to \citet{Sou2009}. In short, we performed aperture photometry using an {\sc idl} implementation of {\sc daophot} \citep{Stetson1987}, and adjusted the aperture sizes to obtain the best results (see Table\,\ref{tab:obslog}). A differential-magnitude light curve was calculated between the target and comparison star. A first-order polynomial was then fitted to the outside-transit data and then subtracted to remove a slow trend present in the differential magnitudes. The times of the start of the exposures were given in JD/UTC in the FITS file headers, and we converted these to times of the midpoints of the exposures in BJD/TDB.

In order to confirm the low scatter of the resulting light curves we performed a second data reduction with completely different methods. We used the \textsc{starlink/autophotom} package \citep{Eaton++99} driven by a custom C-shell script \citep{Me++04mn}, and obtained a light curve with an rms scatter of 414\,ppm for the first night of data. This result agrees with our light curve from {\sc daophot}, once the discretization of the datapoints ({\sc autophotom} quotes instrumental magnitudes to only three decimal places) is taken into account.


\section{Data analysis}
\label{sec:results}

We fitted the WASP-50 data in a similar manner to \citet{Jeremy2012}. \textsc{prism}\footnote[1]{Available from http://www.astro.keele.ac.uk/$\sim$jtr} (Planetary Retrospective Intergrated Starspot Model) was used to model the transit. \textsc{prism} uses a pixellation approach to represent the star and planet on a two-dimensional array in Cartesian coordinates. This makes it possible to model the transit, limb darkening and starspots on the stellar disc simultaneously. Limb darkening was implemented using the standard quadratic law. \textsc{prism} used the six parameters given in Table\,\ref{tab:results} to model the system, where the fractional stellar and planetary radii are defined as the absolute radii scaled by the semimajor axis ($r_{\rm s,p} = R_{\rm s,p}/a$). 

\textsc{gemc}\footnotemark[1] (Genetic Evolution Markov Chain) was used to fit the model to the data. \textsc{gemc} begins by randomly generating parameters for the starting points of $N$ chains, within the user-defined parameter space, and then simultaneously evolves the chains for $X$ generations. At each generation the chains are evaluated for their fitness\footnote[2]{A solution's fitness was found by calculating the $1 / \chi^2$ value.}. The parameters of the fittest member undergo a maximum $\pm 1\%$ perturbation and its fitness is then re-evaluated. If the fitness has improved it is accepted but if the fitness has deteriorated it may or may not be accepted based on a Gaussian probability distribution:

\begin{equation}
P = \exp\left(\frac{\left(\chi^2_{(n-1)} - \chi^2_{(n)}\right)}{2}\right)
\end{equation}

where $(n-1)$ is the previous generational chain and $n$ is the current generational chain being evaluated. The next step is to then evolve the other chains. This is accomplished in a similar way as a genetic algorithm, in that the chain parameters are modified by incorporating the parameters of another chain. But unlike a genetic algorithm where a member is picked by a weighted random number and then the digits of each parameter are crossed over with the digits from a different member, \textsc{gemc} directly perturbs the parameters of each chain in a vector towards the best-fitting chain. The size of this perturbation is between zero and twice the distance to the best-fitting chain, allowing the chain to not only move towards but also to overshoot the position of the best-fitting chain. This continues until all the chains have converged around the optimal solution. Once the burn-in phase has been completed the chains cease communication and begin independent Markov chain runs.

\begin{figure*} \includegraphics[width=0.92\textwidth,angle=0]{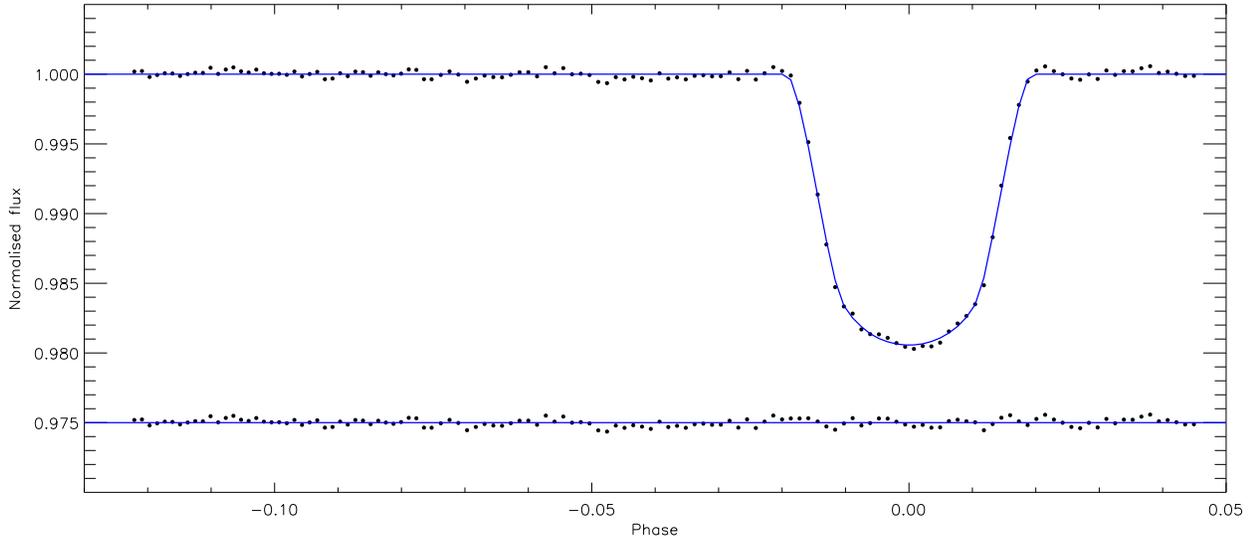}
\caption{\label{fig:lightcurve1} Transit light curve and the best-fitting 
model for 2011/11/20. The residuals are displayed at the base of the figure.} \end{figure*}

\begin{figure*} \includegraphics[width=0.92\textwidth,angle=0]{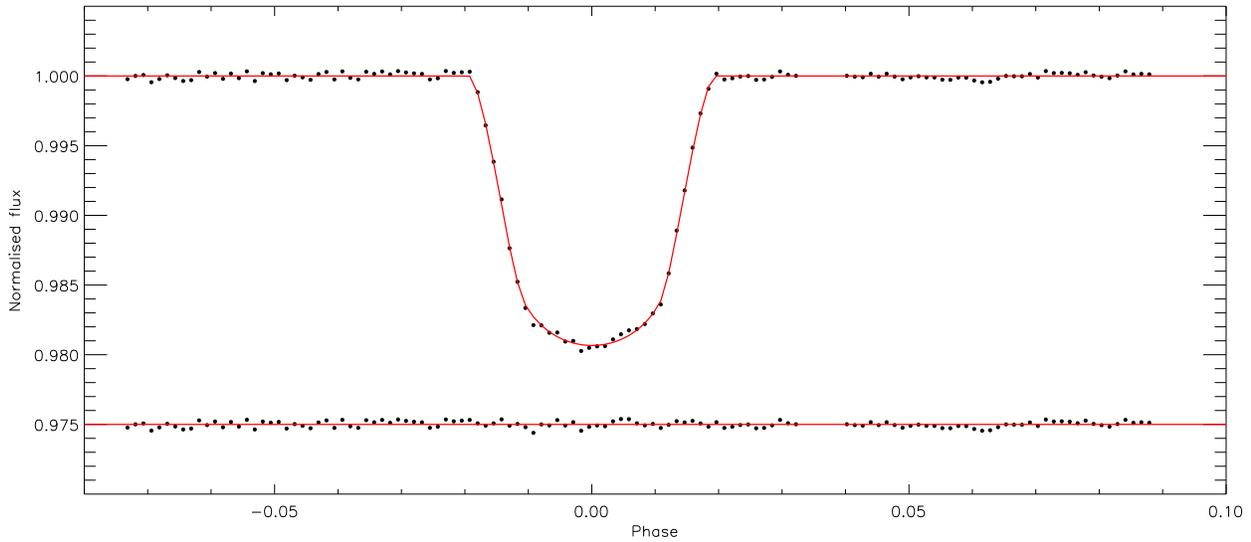}
\caption{\label{fig:lightcurve2} Transit light curve and the best-fitting 
model for 2011/11/24. The residuals are displayed at the base of the figure. 
The gap in the data between phases 0.03 and 0.04 was caused by a technical 
difficulty with the shutter on EFOSC2. This was corrected by a shutdown and 
restart of the instrument. The telescope pointing was unaffected.} \end{figure*}

Because \textsc{gemc} is a hybrid between the Markov Chain Monte Carlo approach and a genetic algorithm, the burn-in phase is relatively short, allowing us to use a large parameter search space. The boundaries of the search space for each parameter are given in Table\,\ref{tab:results}, which also contains the individual results for the two light curves. Table\,\ref{tab:results} also gives the final photometric parameters for the WASP-50 system, which are weighted means of the results from the two individual fits. All errorbars denote 1-$\sigma$ uncertainties. Fig.\,\ref{fig:lightcurve1} and \ref{fig:lightcurve2} compares the light curves to the best-fitting models, including the residuals. The two datasets were modelled individually, and the agreement between the best-fit parameters is exceptionally good. The best-fit limb darkening coefficients are also in good agreement with the theoretically predicted values for WASP-50\,A of $u1 = 0.407$ and $u2= 0.281$ \citep{Claret2004}.

\begin{table*} \centering
\setlength{\tabcolsep}{4pt}
\caption{\label{tab:results} Derived photometric parameters from each lightcurve, plus the interval within which the best fit was searched for using {\sc gemc}.}
\begin{tabular}{lccccc} \hline
Parameter & Symbol & Search interval & 2011/11/20 & 2011/11/24 & Combined photometric parameters \\
\hline
Radius ratio             & $r_p/r_s$   & 0.05 to 0.30      &       0.13710 $\pm$ 0.00049  &        0.13661 $\pm$ 0.00036  & 0.13678 $\pm$ 0.00029 \\
Sum of fractional radii  & $r_s + r_p$ & 0.10 to 0.50      &        0.1552 $\pm$ 0.0018   &        0.1553  $\pm$ 0.0016   &  0.1552 $\pm$ 0.0012  \\
Linear LD coefficient    & $u_1$       & 0.0 to 1.0        &         0.386 $\pm$ 0.068    &        0.385   $\pm$ 0.049    &   0.386 $\pm$ 0.040   \\
Quadratic LD coefficient & $u_2$       & 0.0 to 1.0        &         0.281 $\pm$ 0.099    &        0.279   $\pm$ 0.043    &   0.280 $\pm$ 0.040   \\
Inclination (degrees)    & $i$         & 70.0 to 90.0      &         84.43 $\pm$ 0.17     &        84.45   $\pm$ 0.14     &   84.44 $\pm$ 0.11    \\
Transit epoch (BJD/UTC)  & $T_0$       & $\pm$0.5 in phase & 2455855.78172 $\pm$ 0.000076 & 2455859.691755 $\pm$ 0.000118 &                       \\
\hline \end{tabular} \end{table*}                                                                                                                             

Our data were taken with 120\,s and 150\,s exposures, so we have checked whether these relatively long exposure times affect the derived parameters. For this we modelled the data using the {\sc jktebop} code \citep{Southworth2004}, finding results in good agreement with those from {\sc prism}. We then used {\sc jktebop}'s option to numerically integrate the model over the duration of each exposure whilst finding the best fit \citep{Me11mn}. The final parameters for each light curve altered by only 0.1 to 0.25-$\sigma$, allowing us to conclude that smearing of the transit shape due to long exposure times does not have a significant effect on our results.

To check for correlated `red' noise we used the Monte Carlo and residual-permutation algorithms in \textsc{jktebop} \citep{Me08mn} to assess the uncertainties in the fitted parameters. We found a difference between the two methods of only 0.1\%, and conclude that correlated noise is not present at a significant level in our data.

We also used \textsc{jktebop} to check whether the removal of the slow drift in brightness with a first-order polynomial had any effect on our results. We found that including the polynomial coefficients as fitted parameters caused changes in the other parameters of roughly $0.001$-$\sigma$. We conclude that the detrending process has had no deleterious effect on our results.

\begin{figure*} \includegraphics[width=\textwidth,angle=0]{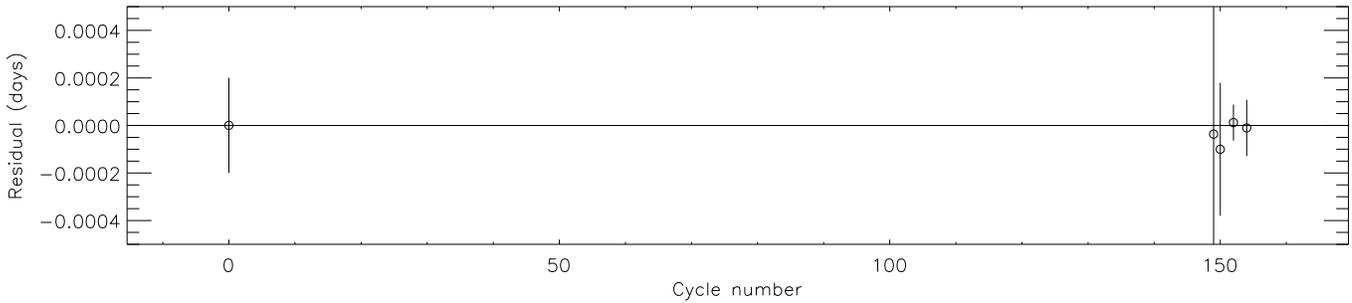} 
\caption{\label{fig:oc} Residuals of the available times of mid-transit 
versus the orbital ephemeris found in this work. The two timings from 
this work are the last two points after cycle number 150.} \end{figure*}

We have collected the available times of mid-transit for WASP-50 from the literature \citep{Gillon2011,Sada2012}. All timings were converted to the BJD/TDB timescale and used to obtain an improved orbital ephemeris:
$$ T_0 = {\rm BJD/TDB} \,\, 2\,455\,558.61237 (20) \, + \, 1.9550938 (13) \times E $$
where $E$ represents the cycle count with respect to the reference epoch and the bracketed quantities represent the uncertainty in the final digit of the preceding number. Fig.\,\ref{fig:oc} and Table\,\ref{tab:minima} show the residuals of these times against the ephemeris. We find no evidence for transit timing variations in the system.

\begin{table} \begin{center}
\caption{\label{tab:minima} Times of minimum light of WASP-50
and their residuals versus the ephemeris derived in this work.
\newline {\bf References:}
(1) \citet{Gillon2011};
(2) \citet{Sada2012}; 
(3) This work.}
\begin{tabular}{l@{\,$\pm$\,}l r r l} \hline
\multicolumn{2}{l}{Time of minimum}   & Cycle  & Residual & Reference \\
\multicolumn{2}{l}{(BJD/TDB $-$ 2400000)} & no.    & (BJD)    &           \\
\hline
55558.61237 & 0.00020 &     0.0 &  0.00000 &  1 \\   
55849.92131 & 0.00060 &   149.0 & -0.00004 &  2 \\   
55851.87634 & 0.00028 &   150.0 & -0.00010 &  2 \\   
55855.78664 & 0.00008 &   152.0 &  0.00001 &  3 \\   
55859.69680 & 0.00012 &   154.0 & -0.00001 &  3 \\   
\hline \end{tabular} \end{center} \end{table}

\subsection{Physical properties of the WASP-50 system}

\begin{table} \centering
\caption{\label{tab:model} Physical properties of the WASP-50 system. 
The equilibrium temperature, \Teq, is for an assumed zero albedo and full heat redistribution. 
\safronov\ is the \citet{Safronov72} number.}
\begin{tabular}{l r@{\,$\pm$\,}c@{\,$\pm$\,}l r@{\,$\pm$\,}l}
\hline 
Parameter & \mcc{This work} & \mc{\citet{Gillon2011}} \\
\hline
$M_{\rm A}$    (\Msun) & 0.861    & 0.052    & 0.023     & \erc{0.892}{0.080}{0.074} \\
$R_{\rm A}$    (\Rsun) & 0.855    & 0.018    & 0.007     & 0.843 & 0.031             \\
$\log g_{\rm A}$ (cgs) & 4.509    & 0.012    & 0.004     & 4.537 & 0.022             \\
$\rho_{\rm A}$ (\psun) & \mcc{$1.376 \pm 0.032$}         & \erc{1.48}{0.10}{0.09}    \\[2pt]
$M_{\rm b}$    (\Mjup) & 1.437    & 0.063    & 0.025     & \erc{1.468}{0.091}{0.086} \\
$R_{\rm b}$    (\Rjup) & 1.138    & 0.024    & 0.010     & 1.153 & 0.048             \\
$g_{\rm b}$     (\mss) & \mcc{$27.50 \pm  0.64$}         & 27.5 & 1.6                \\
$\rho_{\rm b}$ (\pjup) & 0.911    & 0.032    & 0.008     & \erc{0.958}{0.095}{0.082} \\[2pt]
\Teq\              (K) & \mcc{$1410 \pm   26$}           & 1393 & 30                 \\
\safronov\             & 0.0853   & 0.0024   & 0.0007    & \mc{ }                    \\
$a$               (AU) & 0.02913  & 0.00059  & 0.00025   & 0.02945 & 0.00085         \\
Age              (Gyr) & \ermcc{8.1}{6.7}{4.4}{1.5}{1.3} & \mc{ }                    \\
\hline \end{tabular} \end{table}

Once the photometric parameters of WASP-50 had been measured, we moved to the determination of its physical characteristics. We adopted the approach of \citet{Me09mn}, which uses the parameters measured from the light curves and spectra, plus tabulated predictions of several theoretical models. We adopted the values of $i$, $r_p/r_s$ and $r_s + r_p$ from Table\,\ref{tab:results}, and the stellar properties of effective temperature $\Teff = 5400 \pm 100$\,K, metal abundance $\FeH = -0.12 \pm 0.08$ and velocity amplitude $K_{\rm s} = 256.6 \pm 4.4$\ms\ \citep{Gillon2011}.

An initial value of the velocity amplitude of the planet, $K_{\rm p}$, was used to calculate the physical properties of the system using standard formulae and the physical constants listed by \citet{Me11mn}. The mass and \FeH\ of the star were then used as interpolates within tabulated predictions from stellar theoretical models, in order to find the expected \Teff\ and radius. $K_{\rm p}$ was then iteratively refined to find the best agreement between the observed and predicted \Teff, and the light-curve-derived $r_{\rm s}$ and predicted $\frac{R_{\rm s}}{a}$. This was performed for ages ranging from zero age to the terminal-age main sequence, in steps of 0.01\,Gyr. The overall best fit was identified, yielding estimates of the physical properties of the system and the evolutionary age of the star. This procedure was performed separately using five different sets of stellar theoretical models \citep[see][]{Me10mn}, and the spread of values for each output parameter was used to determine a systematic error. Statistical errors were propagated by perturbing each input parameter in turn to quantify the effect on each output parameter. 

The final results of this process have good internal agreement (between the five sets of theoretical models) and are also consistent with those found by \citet{Gillon2011}. The final physical properties are given in Table\,\ref{tab:model} and include separate statistical and systematic errorbars for those parameters with a dependence on the theoretical models. The final statistical errorbar for each parameter is the largest of the individual ones from the solutions using each of the five different stellar models. The systematic errorbar is the largest difference between the mean and the individual values of the parameter from the five solutions. 


\section{Summary and discussion}
\label{sec:Conclusions}

In the pursuit of obtaining accurate properties for transiting extrasolar planetary systems we have achieved photometric precisions of 258 and 211\,ppm in observations of WASP-50, which to our knowledge is a record for ground-based photometry of a point source. Our approach was to heavily defocus the 3.6\,m NTT and to use exposure times of 120--150\,s. We also benefitted from the presence of a good comparison star, at a distance of 2.25\,arcmin from WASP-50 and with similar colours and $r$-band apparent magnitude. The sky brightness was also low, as the moon was below the horizon for most of our observations. 

We reduced our data using two independent pipelines, finding agreement between them. We modelled the light curves using the {\sc prism} and {\sc jktebop} codes, again finding good agreement. From these results and published spectroscopic measurements we have deduced the physical properties of the WASP-50 system to high precision. The properties of the planet WASP-50\,b are now known to within 5\% (mass), 2\% (radius), 4\% (density) and 2\% (surface gravity). This compares to 6\%, 4\%, 9\% and 6\%, respectively, in the discovery paper \citep{Gillon2011}. We also obtained a refined orbital ephemeris. Further improvements in precision could be made in the shorter term by obtaining additional radial velocity measurements, and in the longer term by using sets of stellar models which show a better interagreement on properties of the host star WASP-50\,A.

In our study of WASP-19 \citep{Jeremy2012} we found a modest discrepancy between limb darkening coefficients measured from three datasets taken with the same telescope. This was attributed to the fact that WASP-19\,A is an active star with significant starspot activity, which alters the limb-darkening behaviour of the star \citep{Ballerini2012}. Whilst WASP-50 does show modest chromospheric activity, as judged from emission in the Ca\,II H and K lines \citep{Gillon2011}, starspot anomalies have not been observed in any of the five transit light curves of this system. The limb darkening coefficents found from our two datasets are in excellent agreement (0.02-$\sigma$), supporting the suggestion that starspots affect stellar limb darkening.

\begin{figure} \includegraphics[width=0.46\textwidth,angle=0]{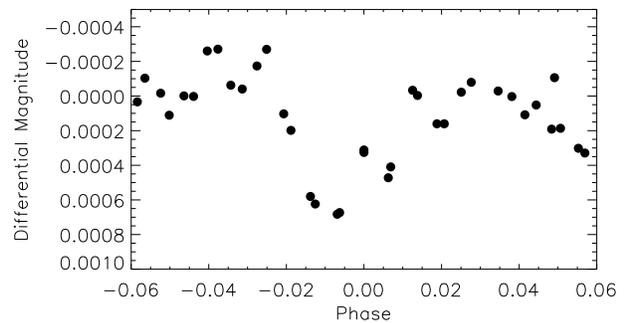} 
\caption{\label{fig:sim} Simulated light curve of a 2\,R$_\oplus$ planet orbiting WASP-50\,A.} \end{figure}

Finally, we have checked if observations of this high precision could be used to characterise transiting super-Earths. We injected a synthetic transit of a 2\,R$_\oplus$ planet in front of WASP-50\,A into the residuals of our best fits from both nights, and binned the data together. This simulated light curve (Fig.\,\ref{fig:sim}) shows a clear transit signature, suggesting that ground-based defocussed photometry of transiting super-Earths is a viable possibility.

\section{Acknowledgements}
\label{sec:Acknow}

We like to thank the anonymous referee for the helpful comments on the manuscript. The reduced light curves presented in this work will be made available at the CDS (http://cdsweb.u-strasbg.fr/) and at both http://www.astro.keele.ac.uk/$\sim$jtr/ and http://www.astro.keele.ac.uk/$\sim$jkt/. This paper used data collected from the New Technology Telescope at the European Southern Observatory, Chile, program number 088.C-0204(A). JTR acknowledges financial support from STFC in the form of a Ph.D.\ studentship. JS acknowledges financial support from STFC in the form of an Advanced Fellowship.


\bibliographystyle{mn_new}


\end{document}